\documentclass[12pt,twoside]{article}
\usepackage{amsmath,amssymb}

\let\lsim\lesssim
\let\gsim\gtrsim

\begin{document}

\title{QCD and Natural Philosophy}
\author{Frank Wilczek\\
\small\it Center for Theoretical Physics\\
\small\it Massachusetts Institute of Technology\\
\small\it Cambridge, MA 02139-4307}
\date{\small MIT-CTP \# 3328 \\[1ex]
Plenary talk at TH2002, Paris, July 2002.}

\maketitle

\markboth{Frank Wilczek}{Natural Philosophy}

\begin{abstract}\noindent QCD sheds considerable light on several of the
most basic features of the natural world, including the origin of
mass, the feebleness of gravity, the extent to which the properties of
matter can be determined conceptually, the
possible utility of the anthropic principle,
and the metatheoretic notions of effectiveness and computability.   I
discuss these applications here.
\end{abstract}

\thispagestyle{empty}
\pagebreak

Contemporary academic or {\it un}natural philosophy is characterized by
extreme insistence on purity of abstract thought.  A joke will
illustrate the point.

A man walks into a bar, takes a seat on the next-to-last stool, and spends
the evening chatting up the empty stool next to him, being
charming and flirtatious, as if there were a beautiful woman in that empty
seat.  The next night, same story.  And the next night, same
story again.  Finally the bartender can't take it any more.  She asks,
``Why do you keep talking to that empty stool as if
there were a beautiful woman in it?''

The man answers, ``I'm a philosopher.  Hume taught us that it's logically
possible that a beautiful woman will suddenly materialize on
that stool, and no one has ever refuted him.   If one does appear,  then
obviously  I'll seem very clever indeed, and  I'll have the inside
track with her.''

``That's ridiculous,'' says the bartender, who happens to be a physicist.
``Plenty of  very attractive women come to this bar all the time.
You're reasonably presentable, and extremely articulate; if you applied
your charm on one of them, you might succeed.''

``I thought about trying that,'' he replies, ``but I couldn't prove it
would work.''

In this vein, there are a lot of very attractive questions that working
physicists normally abandon to the philosophers, or to quasi-scientists
who speculate in reckless disregard of the facts, just because the
questions are broad and qualitative.  We shouldn't.   I'd like in this talk
to take
a stab at a few such questions.  These are questions for which a  physical
theory firmly established only relatively recently, quantum chromodynamics
or QCD, affords us considerable new insight.   I will not waste words
introducing the technical content of QCD, nor defending its validity, as
this
background information is readily available in many places \cite{fw}.

\section{What is the Origin of Mass?}

\subsection{Framing the Question}

That a question makes grammatical sense does not guarantee that it is
answerable,
or even coherent.   In that spirit, let us begin with a critical
examination of the question posed in this Section: What is the origin of mass?

In classical mechanics mass appears as a
primary concept.   It was a very great step for the founders of classical
mechanics to
isolate the scientific concept of mass.   In Newton's laws of motion, mass
appears as an
irreducible, intrinsic property of matter, which relates its manifest response
(acceleration) to an abstract cause (force).    An object without mass
would not know
how to move.  It would not know, from one moment to the next, where in
space it was
supposed to be.  It would be, in a very strong sense, unphysical.   Also
in Newton's law of gravity, the mass of an object
governs the strength of the force it exerts.  One cannot
build up an object that gravitates out of material that does not.  Thus it
is difficult to
imagine, in the Newtonian framework, what could possibly constitute an
``origin of
mass''.    In that framework, mass just is what it is.

Later developments in physics make the concept of mass seem less
irreducible.   The undermining process started in earnest with the theories of
relativity.    The famous equation $E = mc^2$ of special relativity theory,
written
that way, betrays the prejudice that we should express energy in terms of
mass.   But
we can also read it as $m= E/c^2$, which  suggests the possibility of
explaining mass
in terms of energy.   In general relativity the response of matter to
gravity is
independent of mass (equivalence principle), while space-time curvature is
generated directly by energy-momentum, according to
$R_{\mu\nu} - \frac12 g_{\mu\nu} R = \kappa T_{\mu\nu}$, with
$\kappa \equiv 8\pi G_N /c^2$.   Mass appears as a contributing factor to
energy-momentum, but it has no uniquely privileged status.

At an abstract level, mass appears as a label for irreducible
representations of the
Poincar\'{e} group.  Since representations with $m\neq 0$ appear in tensor
products of
$m=0$ representations, it is possible, at least kinematically, to build
massive particles
as composites of massless particles, or of massless particles and fields.

\subsubsection{Lorentz's Dream}

At a much more concrete level, the question of the origin of mass virtually
forced
itself upon physicists' attention in connection with the development of
electron
theory.   Electrons generate electromagnetic fields; these fields have
energy and
therefore inertia.  Indeed,  a classical point-electron is surrounded by an
electric
field varying as
${e}/{r^2}$.   The energy in this field is infinite, due to a divergent
contribution
around $r
\rightarrow 0$.   It was a dream of Lorentz (pursued in evolved forms by many
others, including Poincar\'{e}, Dirac, Wheeler, and Feynman), to account for the
electron's mass entirely in terms of its electromagnetic fields by using a more
refined picture of electrons.  Lorentz hoped that in a correct model of
electrons they
would emerge as extended objects,  and that the energy in the Coulomb field
would
come out finite, and  account for all (or most) of the inertia of electrons.

Later progress in the quantum theory of electrons rendered this program moot by
showing that the charge of an electron,  and therefore of course its associated
electric field,  is intrinsically smeared out by quantum fluctuations in
its position.
Indeed, due to the uncertainty principle the picture of electrons as ideal
point-particles certainly breaks down for distances $r \lsim
\hbar/mc$, the Compton radius.  At momenta $p \gsim \hbar/r$, the velocity
$p/m$ formally becomes of order
$c$, and one cannot regard the electron as a static point source.   If we
cut off the
simple electrostatic calculation at the Compton radius, we find an
electromagnetic
contribution to the electron mass of order
$\delta m \sim \alpha m$, where
$\alpha = e^2/4\pi\hbar c
\approx 1/137$ is the fine structure constant.   In this sense the easily
identifiable
and intuitive electromagnetic contribution to the mass, which Lorentz hoped to
build upon, is small.  To go further, we cannot avoid considering
relativity and
quantum mechanics together.    That means quantum field theory.

\subsubsection{Its Debacle}

In quantum electrodynamics itself, the whole issue of the electromagnetic
contribution to the electron mass becomes quite dodgy, due to
renormalization.

Quantum electrodynamics does not exist nonperturbatively.  One can regulate and
renormalize order-by-order in perturbation theory, but there are strong
arguments
that the series does not converge, or even represent the asymptotic
expansion of a
satisfactory theory.  In a renormalization group analysis, this is because
the effective
coupling blows up logarithmically at short distances, and one cannot remove the
cutoff.  In a lattice regularization, one could not achieve a Lorentz-invariant
limit.\footnote{Actually this blow-up, the famous  Landau pole, arises from
extrapolating the perturbative result beyond its range of validity.   What
one can
deduce simply and rigorously is that the effective coupling does not become
small at
short distances: QED is not asymptotically free.    If there is a fixed
point at finite
coupling, it may be possible to construct a relativistically invariant
limiting theory.
But even if such a theory were to exist, its physical relevance would be quite
dubious, since we know that there's much more to physics than
electrodynamics at
distances so short that the logarithms matter.}   So one cannot strictly
separate the
issue of electromagnetic mass from the unknown physics which ultimately
regularizes
the short-distance singularities of QED.

If we
regard QED as an effective theory, essentially by leaving in an energy cutoff
$\Lambda$, corresponding to a distance cutoff $\hbar c/\Lambda$, we get a
contribution to the mass at short distances going as
$\delta m \propto \alpha m\log (\Lambda/m)$.  Quantum mechanics has changed
the power-law divergence into a logarithm.  As a result, $\delta m$ is a
fractionally
small contribution to the total mass, at least for sub-Planckian
$\Lambda$ (i.e., $\Lambda
\lsim 10^{19}$ GeV).  We know that QED ceases to be a complete description of
physics, or even a well-isolated theory, far below such energies.

In any case, since
the mass renormalization is multiplicative,
an electron field whose bare quanta have bare mass is zero produces physics
quanta with physical
mass zero.   Indeed, the massless version of QED has enhanced symmetry --
chiral symmetry --
which is not destroyed by (perturbative) renormalization.

In short, very little seems to survive from Lorentz's original dream. I've
described
this fiasco in some detail, since it provides an instructive background to
contrast
with our upcoming considerations.

\subsubsection{Upping the Ante}

Quantum field theory changes how we view the
question of the origin of mass dramatically.  As we have seen, it quashes
hope for a simple classical
mechanistic explanation.

At a more profound level, however, quantum field theory makes the
question of the origin of mass seem both better posed and more crucial.
Very few such theories are known to exist nonperturbatively in four
space-time dimensions.   Even if we relax our standards to
include renormalizable quantum field theories, i.e. those which are
reasonably self-contained as effective field theories, we find a very
restricted set.  Such theories, unless they contain many fields and few
symmetries, allow few continuous parameters.  Among those few,
parameters specifying bare masses of the quanta feature prominently.   So
massless theories are even more tightly constrained, and bring us
closer to the ideal of a parameter-free, purely conceptual description of
Nature.

Moreover, theories often acquire  enhanced symmetry
when mass parameters vanish.  In such cases masslessness is ``natural'', in
the sense that it can be stated as an abstract
structural property of the theory, namely its enhanced symmetry.  To the
extent we feel a more symmetric theory is more beautiful,
the massless theories are singled out as more beautiful.   These features
of naturalness and beauty, which
motivate massless theories, survive even if one is not committed to the
technical requirement of renormalizability.

\subsection{Most of the Answer: QCD Lite}

Enough of generalities!  I want now to describe some very specific, and I
think quite beautiful,
insights into the origin of mass in the real world.   We will construct --
following
Nature -- mass without mass, using only $c$ and $\hbar$.

\subsubsection{Introducing QCD Lite}

My central points are most easily made with reference to a slight
idealization of QCD
I call, for reasons that will be obvious, QCD Lite.  It is a  nonabelian gauge
theory based on the gauge group $SU(3)$ coupled to two triplets and two
antitriplets of left-handed fermions, all with zero mass.   Of course I
have in mind
that the gauge group represents color, and that one set of triplet and
antitriplet will
be identified with the quark fields $u_L, u_R$ and the other with
$d_L, d_R$.

Upon demanding renormalizability,\footnote{Or, in physical terms, the
existence of
a relativistically invariant limiting theory.  Or alternatively, radical
decoupling from
an unspecified, gratuitous high-energy cutoff.} this theory appears to contain
precisely one parameter, the coupling $g$.   It is, in units with $\hbar =
c=1$, a pure
number.    I'm ignoring the $\theta$ parameter, which has no physical
content here,
since it can be absorbed into the definition of the quark fields.   Mass
terms for the
gluons are forbidden by gauge invariance.  Mass terms for the quarks are
forbidden
by chiral $SU(2)_L \times SU(2)_R$ flavor symmetry.

\subsubsection{Running Coupling; Dimensional Transmutation}

The coupling constant $g$ that appears in the Lagrangian of QCD Lite, like the
corresponding constant $e$ in QED,  is a dimensionless number (in units with
$\hbar = c = 1$).   Likewise for the fine-structure constant $\alpha_s \equiv
g^2/4\pi$.   But the real situation, when we take into account the effect
of quantum
mechanics, is quite different.   Empty space is a medium, full of virtual
particles, and
responds dynamically to charges placed within it.  It can be polarized, and the
polarization clouds surrounding test charges can shield (screen) or enhance
(antiscreen) their strength.  In other words,  quantum-mechanically the
measured
strength of the coupling depends on the distance scale, or equivalently the
(inverse)
energy scale, at which it is measured: $\alpha_s \rightarrow \alpha_s(Q)$.
This
is a central feature of QCD Lite, and of course of QCD itself.    These
theories predict that the effective coupling gets small at large $Q$, or
equivalently at
short distance.

This behavior displays itself in a remarkably direct and tangible form in
the final
states of electron-positron annihilation.     Hadrons emerging from high-energy
electron-positron annihilation organize themselves into collimated jets.
Usually
there are two jets, but occasionally three.   The theoretical
interpretation is profound
but, given asymptotic freedom, straightforward.  The primary products emerging
from the annihilation are a quark and an antiquark.  They emit soft -- that
is, low
energy-momentum -- radiation copiously, but only rarely hard radiation.
That's a
restatement, in momentum space, of asymptotic freedom.  The soft radiation
materializes as many particles, but these particles inherit their direction
of flow from
the quark or antiquark, and therefore constitute a jet.   In the relatively
rare case
that there is hard radiation, that is to say, emission of an energetic
gluon, the gluon
induces its own independent jet.    All this can be made completely
quantitative.
There are precise predictions for the ratio of three- to two-jet events,
the rare
occurrence of four (or more) jet events, how these ratios change with
energy, angular
dependence, and so forth.  The observations agree with these predictions.
Thus
they provide overwhelming, direct evidence for the most basic elements of the
theory, that is the quark-gluon and gluon-gluon couplings.

Because the coupling runs, we can, within any given version of QCD Lite,
measure
any given {\it numerical\/} value
$a=\alpha_s(Q)$, simply by choosing an appropriate
$Q$.  It appeared, classically, that we had an infinite number of different
versions of
QCD Lite, with different values of the coupling parameter.  In reality the only
difference among all these theories, after they are quantized, is the unit
they choose
for measuring mass.  All dimensionless physical parameters, and in
particular all
mass ratios, are uniquely determined.   We can trade the dimensionless
parameter
$g$ for the unit of mass.   This is the phenomenon of dimensional
transmutation.

Of course the value of the overall energy scale makes a big difference when
we come
to couple QCD  Lite, or of course QCD, to the rest of physics.  Gravity,
for example,
cares very much about the absolute value of masses.   But within QCD Lite
itself, if
we compute any dimensionless quantity whatsoever we will obtain a unique
answer,
independent of any choice of coupling parameter.     Thus, properly
understood, the
value of the QCD coupling constant does not so much govern QCD itself --
within its
own domain, QCD is unique --  but  rather how QCD fits in with the rest of
physics.

\subsubsection{Physical Mass Spectrum -- QCD Lite and Reality}

Now let us consider more concretely how these dynamical phenomena lead us to a
nontrivial hadron spectrum. Looking at the classical equations of  QCD,
one would
expect an attractive force between quarks that varies with the distance as
$g^2/4\pi r^2$,  where $g$ is the coupling constant.  This result is
modified, however, by
the effects of quantum fluctuations.   As we have just discussed, the
omnipresent evanescence of virtual
particles renders empty space into a dynamical medium,  whose response
alters the
force law.   The antiscreening effect of virtual color gluons (asymptotic
freedom), enhances the strength of the attraction, by a factor which grows
with the
distance.     This effect can be captured by defining an effective
coupling, $g(r)$,
that grows with distance.

The attractive interaction among quarks wants to bind
them together;  but the potential energy to be gained by bringing quarks
together
must be weighed against its cost in kinetic energy.    In a more familiar
application,
just this sort of competition between Coulomb attraction and localization
energy is
responsible for the stability and finite size of atoms.   Here
quantum-mechanical
uncertainty implies that quark wave-functions localized in space must contain a
substantial admixture of high momentum components.  For a relativistic
particle, this translates
directly into energy.      If the attraction followed Coulomb's law, with a
small
coupling, the energetic price for staying localized would always outweigh
the profit
from attraction, and the quarks would not form a bound state.
Indeed, the kinetic energy $\hbar c/r$ beats the potential energy
$g^2/4\pi r$.   But the running
coupling of QCD grows with distance, and that tips the balance.   The
quarks finally
get reined in, at distances where $\alpha_s(r)$ becomes large.

We need not rely on heuristic pictures, or wishful thinking, to speculate
about the
mass spectrum of QCD Lite.  It has been calculated by direct numerical
integration of
the fundamental equations, using the techniques of lattice gauge
theory.\footnote{There are significant technical issues around realizing chiral
symmetry in numerical work involving discretization on a  lattice.  Recent
theoretical work appears to have resolved the conceptual issues, but the
numerical
work does not yet fully reflect this progress.   To avoid a fussy
presentation I've
oversimplified by passing over these issues, which do not affect my main
point.}
The results bear a remarkable qualitative and semiquantitative resemblance
to the
observed spectrum of nonstrange hadrons, generally at the 10\% level,
comfortably within known sources of error due to finite size, statistics,
etc. -- and
(anticipating) small quark masses.  Of course, in line with our preceding
discussion,
the overall {\it scale\/} of hadron masses is not determined by the theory.
But all
mass ratios are predicted, with no free parameters, as of course are the
resonance
quantum numbers.

QCD Lite is not the real world, of course.   So although in QCD Lite we get
mass
without mass in the strict sense, to assess how much real-world mass arises
this way
we need to assess how good an approximation QCD Lite is to reality,
quantitatively.
We can do this by adjusting the nonzero values of $m_u$ and
$m_d$ to make the spectrum best fit reality, and then seeing how much they
contributed to the fit.\footnote{Again, there are significant technical
issues here,
especially regarding the role of the strange quark.  Fortunately, the
uncertainties are
numerically small.}   Unlike charges in QED, masses in QCD are soft
perturbations, and we
can calculate a meaningful finite difference between the spectra of these two
theories.  There is also a well-developed alternative approach to
estimating the
contribution of quark masses by exploiting the phenomenology of chiral symmetry
breaking.   Either way, one finds that the quark masses contribute at most
a few per
cent to the masses of protons and neutrons.

Protons and neutrons, in turn, contribute more than 99\% of the mass of
ordinary
matter.  So QCD Lite provides, for our purpose, an excellent description of
reality.
The origin of the bulk of the mass of ordinary matter is well accounted
for, in a
theory based on pure concepts and using no mass parameters -- indeed, no
mass {\it
unit\/} --  at all!

\subsubsection{Comparing with the Old Dream}

While our final result fulfills the fantasy of Lorentz's dream, the details
and the mechanism whereby it is achieved are quite different from what he
imagined.

Obviously we are speaking of hadrons, not electrons, and of QCD, not classical
electrodynamics.  The deepest difference, however, concerns the source and
location
of the energy whereby
$m=E/c^2$ is realized.   In Lorentz's dream, the energy was self-energy,
close to the
location of the point particle.  In QCD Lite the self-mass vanishes.
Paradoxically,
there is a sense in which the self-energy of a quark\footnote{Infinite
self-energy
does {\it not\/} conflict with zero mass.
$E=mc^2$ describes the energy of a particle of mass $m$ when it is at rest;
but of
course, as we know from photons, there can also be energy in massless
particles,
which cannot be brought to rest.} is infinite (confinement), but this is
due to the
spatial extent of its color field, which has a tail extending to infinity,
not to any
short-distance singularity.  To make physical hadrons, quarks and gluons
must be
brought together in such a way that the total color vanishes.  Then there is no
infinite tail of color flux.  The different tails of the individual quarks
and gluons have annulled one another.
But at finite distances the
cancellation is incomplete, because Heisenberg's uncertainty principle
imposes an
energetic cost for keeping color charges precisely localized together.
The bulk of
the mass of the hadrons comes from the residues of these long tails, not from
singularities near point-like color charges.

\subsubsection{Comparing with the Old Debacle}

It is instructive to review, point by point, how the obstructions that
quantum field theory appeared to pose for the ``Origin of Mass'' or
``Mass without Mass'' program are resolved in QCD.

First of all QCD does exist even nonperturbatively, so the issues connected
with renormalization can be analyzed consistently down to rock
bottom.

Second, the massless theory is ``natural'' in the technical sense defined
above.  It displays enhanced symmetry.   The gluons have no mass parameter
because there is no such parameter consistent with local gauge symmetry.
For quarks, there is an additional chiral $SU(2)_L\times SU(2)_R$
symmetry, expanded from the vectorial $SU(2)_{L+R}$, when they massless.

Third, these symmetries, which guarantee zero bare mass for the bare
quanta, do not forbid nonzero mass for physical quanta.  The
gauge symmetry acts trivially on the physical quanta (confinement).  The
chiral symmetry is spontaneously broken: it is a property of the underlying
equations but not of their stable solution (i.e., of their physical
realization).   Spontaneously broken chiral symmetry predicts the existence
of massless
collective modes, the Nambu-Goldstone bosons -- pions -- along with many of
their properties; but it does not forbid mass for the other hadrons --
protons, neutrons, etc.

Fourth, the realistic theory, with small nonzero quark masses, is a {\it
soft\/} perturbation of the massless theory.  It does not require that we
make
any alteration in the renormalization procedure.  Thus it is commensurable
with the massless theory.   We can compare them
unambiguously and quantitatively.

Fifth, and finally, this quantitative comparison shows that the nonzero
quark masses in Nature contribute only a small fraction to the masses of
protons and neutrons.    The chiral symmetry of the massless theory is
approximately valid in reality, and is the basis of a very useful
phenomenology.
The mass ratios of hadrons computed in the idealized massless theory are
very nearly the same as in reality (apart, of course, from the pions).

\subsection{(Many) Remaining Issues}

While the dynamical energy of massless QCD accounts for the bulk of mass in
ordinary matter, it is far from providing the only source of mass in Nature.

Mass terms for quarks and charged leptons appear to violate the electroweak
gauge
symmetry
$SU(2)\times U(1)$.   But gauge symmetry cannot be violated in the fundamental
equations -- that would lead to ghosts and/or nonunitarity, and prevent
construction of a sensible quantum theory.  So these masses must, in a
sense,  have
their ``origin'' in spontaneous symmetry breaking.  That is accomplished,
in the
Standard Model, by having a nonsinglet Higgs field acquire a vacuum expectation
value.    Why this value is so small, compared to the Planck scale, is one
aspect of
what is usually called the hierarchy problem.   Why the couplings of this
field are so
disparate -- especially, what is particularly crucial to the structure of
physical
reality,  why its dimensionless couplings to $e, u, d$ are so tiny (in the
range
~$10^{-5}$ to $10^{-6}$) -- is an aspect of what is usually called the flavor
problem.

Then there are separate problems for generating masses of supersymmetric
particles
(soft breaking parameters, $\mu$ term), for generating the mass of cosmological
`dark matter' (this might be included in the previous item!), for
generating neutrino
masses, and now apparently for generating the mass density of empty space
(cosmological term).

Obviously, many big questions about the origin of mass remain.    But I
think we've
answered a major one, the one that is most important quantitatively for
ordinary matter, beautifully and convincingly.

\section{Why is Gravity Feeble?}

Gravity dominates the large-scale structure of the Universe, but only by
default.   Matter arranges itself to cancel electromagnetism, and the
strong and weak
forces are intrinsically short-ranged.    At a more fundamental level,
gravity is
extravagantly feeble.    Acting between protons, gravitational attraction
is about
$10^{-36}$ times weaker than  electrical repulsion.    Where does this
outlandish
disparity from?  What does it mean?

Feynman wrote
\begin{quote} There's a certain irrationality to any work on [quantum]
gravitation,
so it's hard to explain why you do any of it ....  It is therefore clear
that the problem
we working on is not the correct problem;  the correct problem is  What
determines
the size of gravitation?
\end{quote}

I want to argue that in modern physics it is natural to view the problem of
why gravity is
extravagantly feeble  in a new way -- upside-down and through a distorting lens
compared to its superficial appearance.   When viewed this way, it comes to
seem
much less enigmatic.

First let me quantify the problem.   The mass of ordinary matter is
dominated by
protons (and neutrons), and the force of gravity is proportional to
(mass)$^2$.
From Newton's constant, the proton mass, and fundamental constants we can form
the pure dimensionless number
$$ X~=~  G_N m_p^2 /\hbar c
$$ where $G_N$ is Newton's constant, $m_p$ is the proton mass, $\hbar$ is
Planck's constant, and $c$ is the speed of light.    Substituting the
measured values,
we obtain
$$ X~\approx~ 6\times 10^{-39} ~.
$$ This is what we mean, quantitatively, when we say that gravity is
extravagantly
feeble.

We can interpret $X$ directly in physical terms, too.    Since the proton's
geometrical size $R$ is roughly the same as its Compton radius
$\hbar/m_pc$,  the
gravitational binding energy of a proton is roughly $G_N m_p^2/ R \approx
X m_p
c^2$.   So $X$ is the fractional contribution of gravitational binding
energy to the
proton's rest mass!

\subsubsection{Planck's Hypothesis}

An ultimate goal of physical theory is to explain the world purely
conceptually, with
no parameters at all.   Superficially, this idea seems to run afoul of
dimensional
analysis -- concepts don't have units, but physical quantities do!

There is a sensible
version of this goal, however, that is rapidly becoming conventional
wisdom, despite
falling well short of scientific knowledge.  Soon after he introduced his
constant
$\hbar$, in the course of a phenomenological fit to the black-body radiation
spectrum,  Planck pointed out the possibility of building a system of units
based on
the three fundamental constants $\hbar, c, G_N$.    Indeed, from these
three we can
construct a unit of mass $(\hbar c /G_N)^{1/2}$, a unit of of length
$(\hbar G_N/c^3)^{1/2}$, and a unit of time $(\hbar G_N/c^5)^{1/2}$ --
what we now call the Planck mass, length, and time respectively.

Planck's proposal for a system of units based on fundamental physical
constants was,
when it was made, rather thinly rooted in physics.    But by now there are
profound
reasons to regard $c$,
$\hbar$ and $G$ as  {\it conversion factors\/} rather than  numerical
parameters.
In the special theory of relativity there are symmetries relating space and
time --
and $c$ serves as a conversion factor between the units in which
space-intervals and
time-intervals are measured.   In quantum theory,  the energy of a state is
proportional to the frequency of its oscillations -- and
$\hbar$  is the conversion factor.    Thus $c$ and $\hbar$ appear directly as
measures in the basic laws of these two great theories.    Finally, in
general relativity
theory space-time curvature is proportional to the density of energy -- and
$G_N$
(actually $G_N/ c^4$) is the conversion factor.

If we want to adopt Planck's astonishing hypothesis that we must build up
physics
solely from these three conversion factors, then the enigma of $X$'s
smallness looks
quite different.  We see that the question it poses is not``Why is gravity
so feeble?''
but rather ``Why is the proton's mass so small?"   For according to Planck's
hypothesis, in natural (Planck) units the strength of gravity simply is
what it is, a
primary quantity.  So it can only be the proton's mass that provides the tiny
number
$\sqrt X$.

\subsubsection{Running in Place}

That's a provocative and fruitful way to invert the question, because we
now have
a quite  deep understanding of the origin of the proton's mass,  as I've
reviewed above.

The proton mass is determined, according to the dynamics I've described,
by the
distance at which the running QCD coupling becomes strong.  Let's call this the
QCD-distance.    Our question, ``Why is the proton mass  so small?'' has been
transformed into the question, ``Why is the QCD-distance much larger than the
Planck length?''   To close our circle of ideas, we need to explain how it
is that, if only the Planck
length is truly fundamental,  such a vastly different length arises
naturally.

This last elucidation, profound and beautiful, is worthy of the problem.
It has to do
with how the coupling runs, in detail.   When the QCD coupling is weak,
``running'' is
actually a bit of a misnomer.   Rather, the coupling creeps along like a
wounded
snail.  We can in fact calculate the behavior precisely, following the rules of
quantum field theory, and even test it experimentally, as I mentioned
before.  The
inverse coupling varies logarithmically with distance.   So if we want to
evolve an even moderately small coupling into a coupling of order unity, we
must
let it run between length-scales whose ratio is exponentially large.    So
if the QCD
coupling is even moderately small at the Planck length  it
will reach unity only at a much larger distance.

Numerically, what we predict is that $\alpha_s(l_{\rm Pl.})$ at the Planck
length is
roughly a third to a fourth of  what it is observed to be at
$10^{-15}$~cm; that is, $\alpha_s(l_{\rm Pl.}) \approx 1/30$.   We cannot
measure
$\alpha_s(l_{\rm Pl.})$ directly, of course, but there are good independent
reasons,
having to do with the unification of couplings, to believe that this value
holds in
reality.  It is amusing to note that in terms of the coupling itself, what
we require is
$g_s(l_{\rm Pl.}) \approx 1/2$!  From this modest and seemingly innocuous
numerical hypothesis, which involves neither really big numbers nor
speculative dynamics
beyond what is supported by hard experimental evidence, we have produced a
logical explanation of the tiny value of $X$.

\section{The Reduction of Matter}

The reductionist program, roughly speaking, is to build up the description
of Nature from a few laws that govern the behavior of elementary entities,
and that can't be derived from anything simpler.    This definition is
loose at both ends.

On the output side, because in the
course of our investigations we find that there are important facts about
Nature that we have to give up on predicting.  These are what we -- after
the
fact! -- come to call contingencies.  Three historically important examples
of different sorts are the number of planets in the Solar System, the
precise
moment that a radioactive nucleus will decay, or what the weather will be
in Boston a year from today.   In each of these cases, the scientific
community
at first believed that prediction would be possible.  And in each case, it
was a major advance to realize that there are good fundamental reasons why
it
is not possible.

On the input side, because it is difficult --
perhaps impossible -- ever to prove the {\it non-existence\/} of simpler
principles.   I'll revisit this aspect at the end of the lecture.

Nevertheless, and despite its horrible name, the
reductionist program has been and continues to be both inspiring and
astoundingly successful.   Instead of trying to
refine an arbitrary {\it a priori\/} definition of this program, it is more
edifying to discuss its best fruits.   For beyond
question they do succeed to an astonishing extent in ``reducing'' matter to
a few powerful abstract principles and a small number of parameters.

It is most instructive to build up to our most complete reduction of matter
from some intermediate models that are extremely important and useful in
their own right.

\subsection{Structural Chemistry}

The first such model is obtained by regarding both nuclear masses and the
speed of light as effectively infinite.  We then have the theory of
nonrelativistic electrons in the presence of highly localized, static
sources of positive charge, in multiples of $|e|$, where $e$ is the
electron charge.
We compute the energy as a function of the positions of the source charges,
using the Schr\"odinger equation.  Local
minima define molecules, and we can compute their wavefunctions.  This
model gives a good approximation for the vast subject of structural
chemistry.\footnote{There is an additional fine point, that we must specify
some symmetry rules to take account of quantum statistics for identical
nuclei.}  Its only continuous parameters are
$\hbar, e$, and the electron mass $m_e$.   Since $\hbar$ can be considered
as a conversion factor, and the Bohr radius $\hbar^2/m_e e^2$ just sets the
overall length scale, this is essentially a one-parameter theory.  Indeed,
since one cannot manufacture a dimensionless quantity out of $\hbar, e,
m_e$,
the remaining parameter is really a conversion factor too.   Thus we have a
zero-parameter theory of structural chemistry!

A more accurate version -- QED with sources -- recognizes $c \neq \infty$
and brings in real and virtual photons.   Now there is a pure number in the
game, namely of course the fine structure constant $e^2/4 \pi\hbar c$.
Regarding $\hbar$ and $c$ as conversion factors, and taking into account
that $\hbar / m_e c$ can be used to define the overall length scale, we
arrive at a very accurate one-parameter theory of  structural chemistry
and photochemistry.

\subsection{Matter}

The shortcoming of this theory, of course, is that the matter doesn't move
(except by radiative transitions).  The theory is missing vibrational and
rotational levels, not to mention reactions, diffusion, etc.

To allow motion, we must put in finite values for the nuclear masses.  In
doing this, we open
ourselves up to many, many more input parameters.  At a minimum we need the
masses of the different nuclei and isotopes, and for accuracy in details
we need their spins and magnetic moments, etc.  At this level, all of them
must be taken from experiment.  There are several hundreds of parameters.
Nevertheless, because they are highly overconstrained by hundreds of
thousands of measurements, the resulting theory is extremely useful.   It is
the foundation of practical quantum chemistry.

The next step in reduction is the one supplied by QCD.   QCD assures us
that all the nuclear parameters can be calculated crudely -- perhaps at the
10\% level --  using QCD Lite.  To do this, we need only {\it one\/}
parameter beyond $\hbar$ and $c$, as I've already discussed.  That
parameter can be taken as  the coupling
$g_s$, normalized at some appropriate energy scale (say $10^5 m_e c^2$), or
equivalently the energy scale $\Lambda_{\rm QCD}$, which is roughly
speaking the scale at which the coupling becomes large.   To do a more
accurate job, we also need to introduce the light quark masses $m_u, m_d$.

Our faith in QCD is not based primarily on its practical ability to
reproduce nuclear parameters.   The best quantitative tests of the theory
come from an
entirely different domain of behavior, scattering at ultra-high energy,
where the quark and gluon degrees of freedom and their couplings are clearly
revealed.  We rely on the successful outcome of these tests, and (so far)
rather crude results from massive numerical calculations of proton structure
and the hadron spectrum for empirical support.  This support is highly
leveraged by the theory's being rigidly principled.  It embodies
the
basic principles of relativity, quantum mechanics, and gauge symmetry.
Only a few parameters appear in the most general consistent
embodiment of these principles.  They are just the ones we
use!\footnote{There is also the notorious $\theta$ parameter, a can of
worms  I don't want to
dive into just now.}

So in practice the ``reduction'' we offer the chemists, or for that matter
the nuclear physicists,  is not of much use to them.  But it is
there, as a matter of principle.

At this point our best available reduction of ordinary matter is in place.
A fundamental theory that we believe offers an extremely
complete and accurate set of equations governing the structure and behavior of
matter in ordinary circumstances, with a very liberal definition of ``ordinary'',
requires the parameters
$\hbar, c, e, m_e, \Lambda_{\rm QCD}, m_u, m_d$.  Among these $\hbar$ and $c$
can be regarded as conversion factors, $m_u$ and $m_d$ are relatively
minor players, and one combination just sets an overall length scale.  Thus
we have a rough reduction of matter involving two parameters, and an
accurate one involving four.   Not bad!

\subsection{Astrophysics}

In astrophysics, as opposed to condensed matter physics, chemistry, and
biology, gravity and the weak interactions play important roles.  Their
intrinsic weakness is compensated by the availability of vast amounts of
matter and vast amounts of time, respectively.  To treat these additional
interactions we must introduce two additional parameters, the Newton
constant $G_N$ and the Fermi constant $G_F$.  Altogether, then, we now have
$\hbar, c, e, m_e, \Lambda_{\rm QCD}, m_u, m_d, G_N, G_F$.   Referring back to
our discussion of Planck's hypothesis, we are inclined to regard $\hbar,
c$ and $G_N$ as conversion factors.  Also, we derive both $e$ and
$\Lambda_{\rm QCD}$ from the value of the unified coupling at the Planck scale.
So
we are left with this ``reasonable'' number ($g_{\rm unified} \approx 1/2$)
and four ``unreasonable'' numbers: $m_e, m_u, m_d$, and $G_F^{-\frac{1}{2}}$,
that specify various masses in Planck units.

The absurd smallness of these numbers, and of the dimensionless
ratio\footnote{It is
important to note that the dimensionless form of this ratio does not
involve $G_N$.  Its absurd smallness is {\it not\/} an artifact of Planck
units.}
$m_e (G_F/\hbar c)^{\frac{1}{2}}$, define
important unmet challenges for theoretical physics.  They are salient
aspects of what are called the hierarchy and flavor problems, respectively.

Though wonderful challenges remain, the success of the reductionist program
is already most impressive.  An accurate description of matter, including
astrophysics, is built up from a few abstract principles and five parameters.

\subsection{Cosmology and Quantitative Anthropics}

Having come this far it would be inhuman not to take the discussion to its
logical climax in cosmology.    That brings in
many unsettled questions that go far beyond the scope of this lecture.
I'll restrict myself to a very brief discussion of a few issues closely
tied to the
foregoing considerations.  I'll assume as given the emerging ``standard
model of cosmology'', based on small perturbations around
a spatially flat Friedmann-Robertson-Walker model.

To specify quantitatively the standard model of cosmology we must
add a few more parameters to our world-model.   These include at least the baryon
number density
$\rho_b$, the dark matter density $\rho_d$, the overall amplitude $A$ of a
scale-invariant spectrum of
primordial fluctuations (assumed adiabatic and Gaussian), and the
cosmological term $\Lambda$.   Further work may uncover the need for
additional parameters, for example if the Universe is not quite flat, or if
the spectrum of perturbations is not quite scale-invariant, but for now the
four I mentioned appear to form an adequate, minimal set.
If we accept that cosmology is ``reduced'' to general principles and four
continuous parameters, beyond the ones we use to describe matter and
astrophysics, the question arises how far this achievement fulfills the
reductionist program.   I think that most physicists do not feel entirely
satisfied
with it.  That is because the parameters in cosmology, given in this way,
do not appear directly in the description of the behavior of simple
fundamental
entities.  Rather they appear as the description of  average properties of
macroscopic ({\it very\/} macroscopic!) agglomerations.   We would like to
calculate  these parameters in terms of different ones more related to the
fundamental laws of physics in the early Universe.   There are rough ideas
about how we might do this in the cases of $\rho_b$ (B and CP violation)
and $\rho_d$ (axion and/or lightest supersymmetric particle
production), even rougher ideas in the case of $A$ (inflationary
potentials), and various tenuous speculations regarding $\Lambda$.  There
is a very
good chance, I think, that the constitution of the dark matter will detected
experimentally in the near future.  Then we'll have a much
better chance to understand at least $\rho_d$ properly!

Taken this far the status of the reductionist program in cosmology, though
much less fully developed, is not very different in character from the
preceding discussions of matter and astrophysics.  But there are
specifically cosmological issues, too.

Although it is usually passed over in silence, I think it is very important
philosophically, and deserves to be emphasized, that our standard cosmology
is {\it radically\/} modest in its predictive ambitions.   It consigns
almost everything about the world as we find it to contingency.  That
includes
not only the aforementioned question of the number of planets in the Solar
System, but more generally every specific fact about every specific object
or
group of objects in the Universe, apart from a few large-scale statistical
regularities.   Indeed, specific structures are supposed to evolve from the
primordial perturbations, and these are only characterized statistically.
In inflationary models these perturbations arise as quantum fluctuations,
and
their essentially statistical character is a consequence of the  laws of
quantum mechanics.

This unavoidably suggests the question whether we might find ourselves
forced to become even more radically modest.\footnote{A few outliers hope to
move back the other way.  One can only wish them luck!}  Let us suppose
for the sake of argument the best possible case, that we had in hand the
fundamental equations of physics.  Some of my colleagues think they do, or
soon will.   Even then we have to face the question of what principle
determines the solution of these equations that describes the observed
Universe.  Let me again suppose for the sake of argument the best possible
case,
that there is some principle that singles out a unique acceptable solution.
Even then there is a question we have to face: If the solution is
inhomogeneous, what determines our location within it?

Essentially inhomogeneous solutions of fundamental equations are by no
means a contrived or remote possibility.  They are at the heart of all the
brane-world constructions that have received much attention recently; they
are characteristic of eternal inflation models; they arise very
naturally in axion physics and its generalizations \cite{fw}.   On a less
cosmic scale, closer to home, they are what we see all around us -- and of
course,
each of us {\it is\/} a highly structured inhomogeneity.

The standard cosmological model is yet another case, and supplies an
especially clear parable.   As we have just discussed, the laws of
reductionist
physics do not suffice to tell us about the specific properties of the Sun,
or of Earth.   Indeed, there are many roughly similar but significantly
different
stars and planets elsewhere in the same Universe.  On the other hand, we can
aspire to a rational, and even to some extent quantitative, ``derivation''
of
the parameters of the Sun and Earth based on fundamental laws, if we define
them not by pointing to them as specific objects  -- that
obviates any derivation -- but rather by characterizing broad aspects of
their behavior.   In principle any behavior will do, but possibly the
most important and certainly the most discussed is their role in supporting
the existence of intelligent observers, the so-called anthropic principle.
There are many peculiarities of the Sun and Earth that can be explained
this way.   A crude example is that the mass of the Sun could not be
much bigger or much smaller than it actually is because it
would burn out too fast or not radiate sufficient energy,  respectively.

Now if the Universe as we now know it constitutes, like the Solar System,
an inhomogeneity within some larger structure, what might be a sign of
it?\footnote{Of course, to make it interesting I'm assuming that we can't
actually look beyond, due to the vastness of the homogeneous region we're
in, or
the limitations of our probes.}  If the parameters of fundamental physics
crucial to life -- just the ones we've been discussing! -- vary from place
to
place, and most places are uninhabitable, there would be a signature to
expect.  We should expect to find that some of these parameters
appear very peculiar -- highly nongeneric -- from the point of view of
fundamental theory, and that relatively small changes in their values would
preclude the existence of intelligent observers.   Weinberg has made a case
that the value of the cosmological term $\Lambda$ fits this description; and
I'm inclined to think that
$(m_u - m_d)/\Lambda_{\rm QCD}$ and several other combinations of the small
number of ingredients in our reduced description of matter and
astrophysics do too.   A fascinating set of questions is suggested here,
that deserves careful attention \cite{hogan}.

\section{Metatheory}

QCD is unique among physical theories in its combination of logical closure
and empirical success.  So it is appropriate to consider it from the
outside,
as an object of metatheory.

\subsection{The Unreasonable Effectiveness of QCD}

There is a sense in which QCD is better than it  has to be.  It is not a
complete theory of the world.  Even within the standard model
quarks have additional interactions, beyond QCD, that become significant --
indeed, problematic -- at short distances.  So there is no logical
requirement
that QCD, extrapolated on its own down to infinitely short distances,
should exist as a fully consistent relativistic quantum field theory.   But
it does.  Is
this just a happy coincidence,  or does it point to something deeper?
(Candidate idea: There must be fully consistent quantum field theories,
because
such theories are the ultimate description of Nature.  A few years ago this
might have been string-blasphemy; now it might be acceptable as
duality.)

Although it's somewhat off the point, I'll take the opportunity to mention
that this over-effectiveness reminds me of another apparently quite
unrelated case, the cosmic censorship hypothesis in general relativity.  We
know that classical general relativity is not a complete theory, and
there's no
logical reason why it should hide its defects from view.  Yet in many
cases, and perhaps generically, it does seem that singularities get hidden
behind
horizons.

\subsection{The Unreasonable Ineffectiveness of QCD}

There some aspects of QCD I find deeply troubling -- though  I'm not sure if  I
should!

I find it disturbing that it takes vast computer resources, and careful
limiting procedures, to simulate the mass and properties of a proton with
decent
accuracy.  And for real-time dynamics, like scattering, the situation
appears pretty hopeless.
Nature, of course, gets such results fast and effortlessly.  But how, if
not through some
kind of computation, or a process we can mimic by computation?

Does this suggest that there are much more powerful forms of
computation that we might aspire to tap into?  Does it connect to the
emerging theory of quantum computers?  These musings suggest some concrete
challenges: Could a quantum computer calculate QCD processes efficiently?
Could it defeat the sign problem, that plagues all existing algorithms
with dynamical fermions?  Could it do real-time dynamics, which is beyond
the reach of existing, essentially Euclidean, methods?

Or, failing all that,
does it suggest some limitation to the universality of computation?

Deeply related to this is another thing I find disturbing.  If you go to a
serious mathematics book and study the rigorous construction of the real
number system, you will find it is quite hard work and cumbersome.  QCD,
and for that matter the great bulk of physics starting with classical
Newtonian mechanics, has been built on this foundation.  In practice, it
functions quite smoothly.  It would be satisfying, though, to have
a ``more reduced'' description, based on more primitive, essentially
discrete structures.  Fredkin and recently Wolfram have speculated at
length along
these lines.   I don't think they've got very far, and the difficulties
facing such a program are immense.  But it's an interesting issue.

\subsubsection*{Acknowledgments}
This work is supported in part by funds provided by the U.S. Department of
Energy (D.O.E.) under cooperative research agreement \#DF-FC02-94ER40818.  
I'd like to thank the organizers for putting together a varied and fascinating
program, and bringing it all to Paris. And I'd like
specifically to thank Dominique Schiff for encouraging me to give a
``philosophical'' talk, which I had fun doing.  She shouldn't be held
responsible for my excesses, though.

\end{document}